\documentclass[prb,aps,twocolumn, superscriptaddress]{revtex4-2} 

\usepackage{etoolbox}
\makeatletter
\patchcmd{\@author@present@script}
  {\gdef\comma@space{\textsuperscript{,\,}}}
  {\gdef\comma@space{\textsuperscript{\,}}}
  {}{}
\makeatother

\usepackage{graphicx}
\usepackage{dcolumn}
\usepackage{bm}
\usepackage{multirow}

\usepackage{centernot}

\usepackage[normalem]{ulem}
\usepackage{cancel}

\usepackage{amsmath}
\usepackage{amsthm}
\usepackage{amstext}
\usepackage{amssymb}
\usepackage{mathrsfs}
\usepackage{amsfonts}
\usepackage{amsbsy} 
\usepackage{tensor}
\usepackage{physics}

\usepackage{wasysym}

\usepackage{latexsym}
\usepackage[american]{babel}
\usepackage{bbm}
\usepackage[colorlinks=true, citecolor=blue!90!black, linkcolor=blue!90!black, linktocpage=true, urlcolor=red!70!black]{hyperref}

\newcommand*{\figref}[2][]{%
  \hyperref[{#2}]{%
    \ref*{#2}%
    \ifx\\#1\\%
    \else
      \,#1%
    \fi
  }%
}

\usepackage[all,matrix,cmtip]{xy}

\usepackage{csquotes}
\MakeOuterQuote{"}

\usepackage{color}
\definecolor{red}{rgb}{1,0,0}
\definecolor{blue}{rgb}{0,0,1}
\definecolor{dblue}{rgb}{0,0,0.4}
\definecolor{green}{rgb}{0,1,0}
\definecolor{black}{rgb}{0,0,0}
\definecolor{white}{rgb}{1,1,1}
\definecolor{niceBlue}{RGB}{20,10,237}

\usepackage[dvipsnames]{xcolor}

\definecolor{brn}{rgb}{.8,.4,.0}
\definecolor{redo}{rgb}{1,.5,.0}
\definecolor{ddgrn}{rgb}{0,0.4,0}
\definecolor{dgrn}{rgb}{0,0.55,0}
\definecolor{dbl}{rgb}{0,0,0.5}

\usepackage[bbgreekl]{mathbbol}

\newcommand{\p}[1]{\prime\,}

\newcommand{\ee}{\hspace{1pt}\mathrm{e}}
\renewcommand{\dd}{\hspace{1pt}\mathrm{d}}

\newcommand{\bpm}{\begin{pmatrix}}
\newcommand{\epm}{\end{pmatrix}}
\newcommand{\bmm}{\begin{matrix}}
\newcommand{\emm}{\end{matrix}}

\usepackage{euscript}

\newcommand{\eps}{\epsilon}

\newcommand{\Ga}{\Gamma}

\usepackage{tikz}
\usepackage{tikz-cd}
\usetikzlibrary{arrows}
\usetikzlibrary{intersections}
\usetikzlibrary{shapes.geometric}
\usetikzlibrary{decorations.pathmorphing, patterns,shapes}
\usetikzlibrary{decorations.markings}

\usepackage{enumerate}

\usepackage{amsthm}

\theoremstyle{definition}

\usetikzlibrary{calc}

\usepackage{slashed}

\usepackage{booktabs}

\newcommand{\ie}{\begin{equation}\begin{aligned}[]}
\newcommand{\fe}{\end{aligned}\end{equation}}

\usepackage{comment}
\usepackage{mathtools}

\renewcommand{\k}{\mathbf{k}}

\renewcommand{\p}{\mathbf{p}}
\renewcommand{\r}{\mathbf{r}}

\newcommand{\mb}[1]{\mathbf{#1}}
\usepackage{braket}

\renewcommand{\bar}[1]{\overline{#1}}

\usepackage[normalem]{ulem}
\usepackage{subcaption}

\begin{document}

\title{
Shape of Wigner Crystals and Hole Self-Doping in a Mexican-Hat Dispersion
}

\author{Minho Luke Kim}
\affiliation{Department of Physics, Massachusetts Institute of Technology}

\author{Xiao-Gang Wen}
\affiliation{Department of Physics, Massachusetts Institute of Technology}

\begin{abstract}

We study Wigner crystals (WCs) induced by a strong Coulomb interaction from
the ring-like Fermi surface of a Mexican-hat dispersion $\epsilon_k=
c_2k^2+c_4k^4$. We design orbital shape in order to minimize the
energy of the WC, and find that a low ground-state energy requires an orbital
shape with a depletion of electrons near $k=0$. To capture the Coulomb-induced
correlations, we include a Jastrow factor as well as a factor describing the
correlation between electrons and doped vacancies. Using variational Monte
Carlo calculations, we calibrate the effective band parameters $c_2$ and $c_4$ to
reproduce the two transitions observed experimentally as the electron density
is lowered: from a spin-valley-polarized Fermi liquid with a disk-like Fermi
surface, to one with a ring-like Fermi surface, and finally to a WC. We find
that, even with an optimized orbital shape that depletes electrons near $k=0$,
a WC with hole self-doping near $k=0$ can still be energetically favorable
near the WC transition, provided that the electron-vacancy correlation is
included. We also estimate the dispersion of the doped hole.

\end{abstract}

\maketitle

\makeatletter 
\def\l@subsection#1#2{}
\makeatother 


\textit{Introduction}---The Wigner crystal (WC)~\cite{W3446} is an archetypal
phase of matter that emerges when the repulsive Coulomb interaction is
sufficiently strong, or equivalently when the electron density is sufficiently
low. From an energetic point of view, Wigner crystallization occurs when the
potential energy dominates the kinetic energy, causing the electrons to
organize into a crystal in order to remain far apart from one another.

For an ordinary two-dimensional electron gas (2DEG), the transition between a
WC and a Fermi liquid (FL) has been studied
extensively~\cite{TC8905, AB0288, DN09102}, including in the presence of Berry
curvature~\cite{JS250722121} or spin-orbit coupling~\cite{PE13092307}.
Nevertheless, the possibility of competing phases at intermediate interaction
strength remains an open question, even for a conventional 2DEG~\cite{SK0470,
PD0877, HD191007152, KK230913121, CC0002366, KP0331, FW0406600, NP0545}. A
variety of candidate states, ranging from charge-density waves to self-doped
crystals, have been proposed as possible intermediate phases between the FL
and the WC.

In recent years, Wigner crystallization has again come to the forefront in
flat-band systems, where several experiments~\cite{LW210110599, LJ230917436,
TY231211632, ZF240205456, GC251012009} have reported direct evidence for
electron crystallization. Since these systems can also exhibit quantum Hall
and anomalous Hall phenomena, the possibility of anomalous Hall
crystals~\cite{HA8657, KS8656, DS231103445, DP231105568, ZZ231104217,
SP240305522, DS240307873, KB231211617, YB240713770, TD240304196, RF250821000,
TD240906775, ZZ241104174} has become an important topic of discussion. One
notable example is rhombohedral multilayer graphene (RMG), which has been
shown to host a variety of competing phases, including quantum anomalous Hall
states~\cite{LJ230917436, LJ240810203, WY240810133} and chiral
superconductivity~\cite{HJ240815233, QL250405129}, in the vicinity of the
Wigner-crystal and Fermi-liquid regimes.

In addition to quantum Hall physics, a recent experiment on 4- and 6-layer
rhombohedral graphene reported signatures of a possible metallic Wigner
crystal (mWC) state~\cite{HJ260400113} whose conductance appears to be
carried by holes. Although the microscopic mechanism responsible for this mWC
state remains unclear, various candidate states near the crystallization
transition have been explored. In multilayer graphene systems, where quantum
geometry can play an important role, several toy models~\cite{SP240305522,
DS231103445, TD240304196, ZC240217867, ML251223082, JS250722121} have been
developed, most recently including the $\lambda$-jellium
model~\cite{SP250312704, DV260400114, FD260409820}.

\begin{figure}
    \centering
    \includegraphics[height=1.3in]{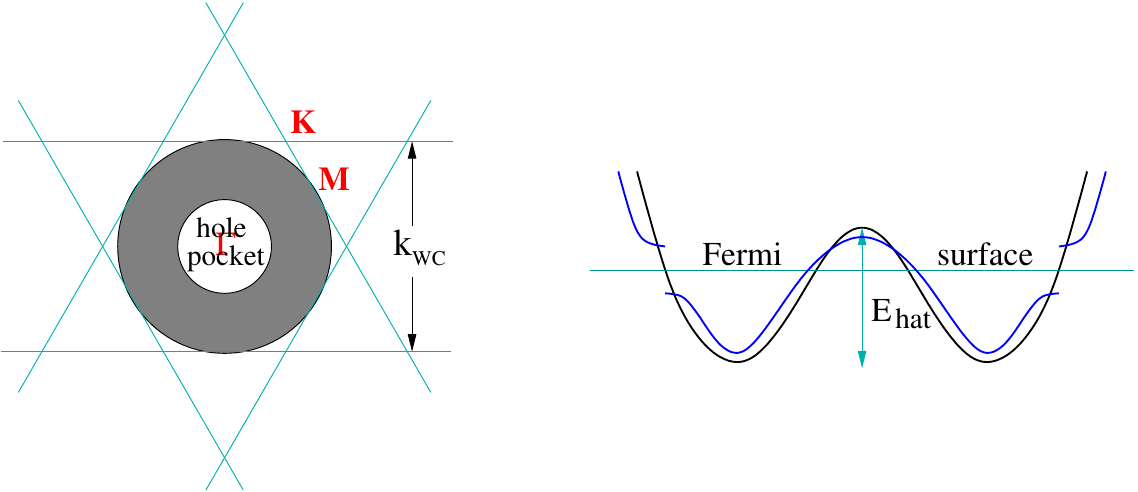}
    \caption{ \textbf{(Left)} In the weak-interaction limit, an induced CDW
may favor a lattice constant that matches the outer edge of the ring-like
Fermi surface associated with the Mexican-hat dispersion, thereby gapping out
the outer Fermi surface. The inner Fermi surface remains, giving rise to an
mWC state.  The shaded region in momentum space denotes the occupied
electronic states, and $k_{\rm WC}$ is the momentum transfer generated by the
WC potential.  \textbf{(Right)} Schematic Mexican-hat dispersion and the
gapping of the outer Fermi surface by the WC potential.  } \label{mCDW}
\end{figure}

In this paper, we adopt a more microscopic viewpoint for the transition from
the ring-like Fermi surface of a Mexican-hat dispersion to a WC (see
Fig.~\ref{mCDW}), and emphasize the role of the strong Coulomb interaction. To
understand the neighboring phases and possible instabilities of a WC, it is
crucial to determine the \emph{shape} of the WC, namely the shape of the
localized electronic wavefunction beyond the simple Gaussian wave-packet WC.
In this work, we focus on this effect while neglecting quantum geometry,
trigonal warping, and other band-topological effects. After systematically
constructing the single-particle orbitals for the commensurate crystals to
accommodate the dispersion, we create a hole in the WC and verify that a
$\Gamma$-hole of the Wigner crystal Brillouin zone can be energetically
competitive against the commensurate crystal, which then could be even further
improved when electron density relaxations around a vacancy (polaron) is
incorporated from a beyond-Jastrow correlation factor. To our knowledge, this
is the first many-body calculation of the these phases in a
Mexican-hat dispersion. Previous related studies~\cite{DV260400114,
FD260409820} instead employed Hartree-Fock calculations for microscopic
models.

\textit{Terminology and summary of result}---For a WC arising from a ring-like
Fermi surface, and in the presence of strong Coulomb repulsion, we expect the
simple Gaussian wavepacket WC (gWC) will be modified so as to reduce its
small-$\k$ weight and thereby lower the kinetic energy, while preserving the
crystalline structure. One way to directly work with the wavepacket is by
applying a high-pass filter to the Gaussian wavepacket and remove
contributions from small $\k$ sectors. We call this modified wavepacket state
the annular WC (aWC), which incorporates the Mexican-hat dispersion by
suppressing small-$\lvert \k \rvert$ components.

A different way to incorporate the depletion of electrons near small $\k$ is
to choose the single-particle wavefunctions as eigenstates in a WC potential
with dispersion $\epsilon_{\k}=c_2 k^2+c_4 k^4$. The many-body ansatz is then
obtained by forming a Slater determinant from the single-particle orbitals in
the \emph{lowest band}. We refer to the resulting many-body state as a
charge-density-wave (CDW) state. This CDW wavefunction is smoothly connected
to the spin-valley-polarized quarter Fermi liquid (QFL) wavefunction. All of
these ans{\"a}tze will then be supplemented by a universal Jastrow factor to
incorporate electron correlations.

\begin{figure}
	\centering
	\includegraphics[width=\linewidth]{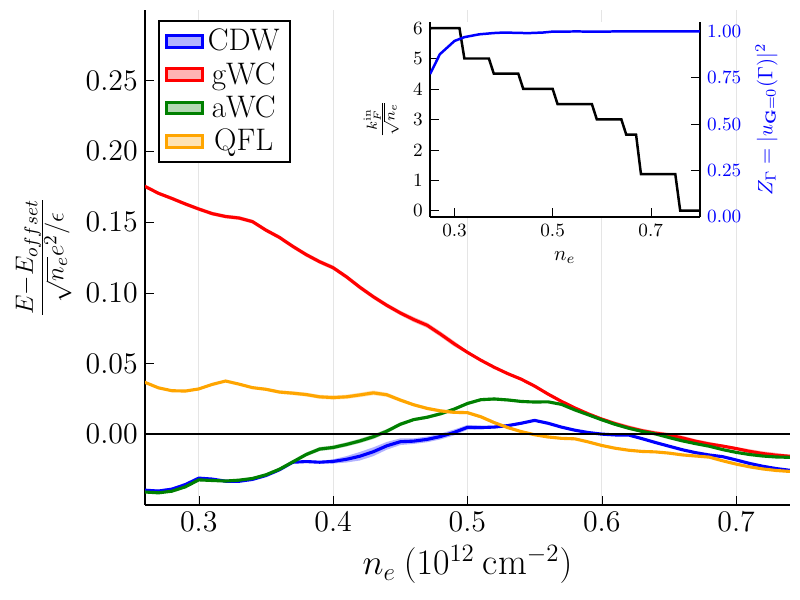}
	\caption{Energies of the commensurate crystal ans{\"a}tze compared to
the QFL ansatz. The CDW ansatz and aWC ansatz are much more energetically
favorable relative to the standard gWC ansatz as it accommodates the Mexican
hat dispersion, with CDW performing better than aWC at intermediate densities
while converging at low densities. Range of errors are given in the ribbon. The density-dependent offset function $E_{offset} = (-3.23 + 2.71 \bar{n}_e - 1.60 \bar{n}_e^ 2) \frac{e^2\sqrt{n_e}}{\epsilon} $ is applied for clearer presentation of energy differences between states where $\bar{n}_e = \frac{n_e}{10^{12} \textrm{cm}^{-2}}$.
\textbf{Inset}: Black - optimized radius of the inner Fermi surface for QFL in
units of $\sqrt{n_e}$, showing Lifshitz transition around $n_e=0.7 \times
10^{12}$cm$^{-2}$. Blue - $Z_\Gamma = \lvert u_{\mb G = 0} (\Gamma) \rvert^2$
by density, which is the contribution of the $\mb G = 0$ element of the Bloch
orbital at $\Ga$. Reciprocal-vector mixing is small for $n_e \geq 0.3 \times
10^{12}$cm$^{-2}$. } \label{fig:WCCDW}
\end{figure}

We choose a dielectric constant $\eps=10$, $c_2=-190~\mathrm{meV}\,
\mathrm{nm}^2$, and $c_4=1000~\mathrm{meV}\,\mathrm{nm}^{4}$. This gives a
Mexican-hat depth $E_\text{hat}=\frac{c_2^2}{4 c_4} = 9.0$~meV in
Fig.~\ref{mCDW}, and a Coulomb energy $e^2\sqrt{n_e}/\eps = 10.2$~meV for $n_e
= 0.5 \times 10^{12}~\mathrm{cm}^{-2}$.
These parameters are chosen so that when the optimized variational energies of
the commensurate ans{\"a}tze and the QFL are compared over the density range
$n_e=0.25$--$0.75\times 10^{12}\,\mathrm{cm}^{-2}$, the Lifshitz transition
occurs around $0.7\times 10^{12}\,\mathrm{cm}^{-2}$ and the crystal state wins
over the QFL at approximately $0.5\times 10^{12}\,\mathrm{cm}^{-2}$, which are
the comparable with the experimental values (see Fig.~\ref{fig:WCCDW}).  The
fact that the aWC and the CDW have much lower energies than the gWC indicates
the importance of including electron depletion near $\k=0$.  A related
variational Monte Carlo (VMC) study for a system with quartic dispersion was
recently carried out in Refs.~\cite{KW240918067, KW250718582}, and here we
employ similar techniques.

In the weak-interaction limit, the CDW induced from a ring-like Fermi surface
is naturally metallic, with a hole-like Fermi surface around $\k=0$ (see
Fig.~\ref{mCDW}). However, the transition from the QFL to the WC is first
order. Thus, once the WC forms, the interactions are already strong, and the
metallic phase in Fig.~\ref{mCDW} may be bypassed.

In the strong-interaction limit, a doped hole in the WC costs a Coulomb energy
of order $e^2\sqrt{n_e}/\eps$, while removing an electron from the $\k=0$
orbital lowers the kinetic energy. It is not clear whether the gain in kinetic
energy can outweigh the cost in Coulomb energy and drive a hole-self-doping
instability of the WC.

\begin{figure}
	\begin{subfigure}{\linewidth}
	\centering
	\includegraphics[width=\linewidth]{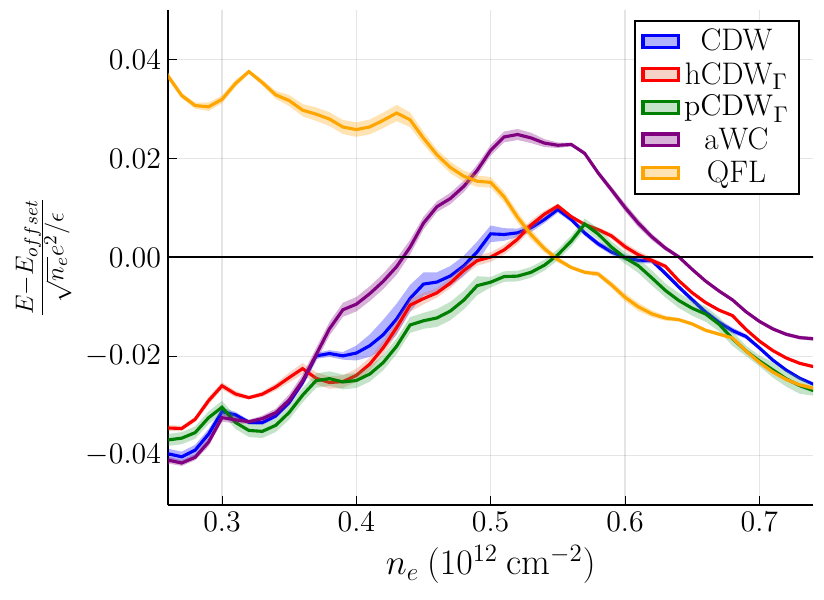}
    \end{subfigure}
	\\
	\begin{subfigure}{\linewidth}
		\centering
		\includegraphics[width=\linewidth]{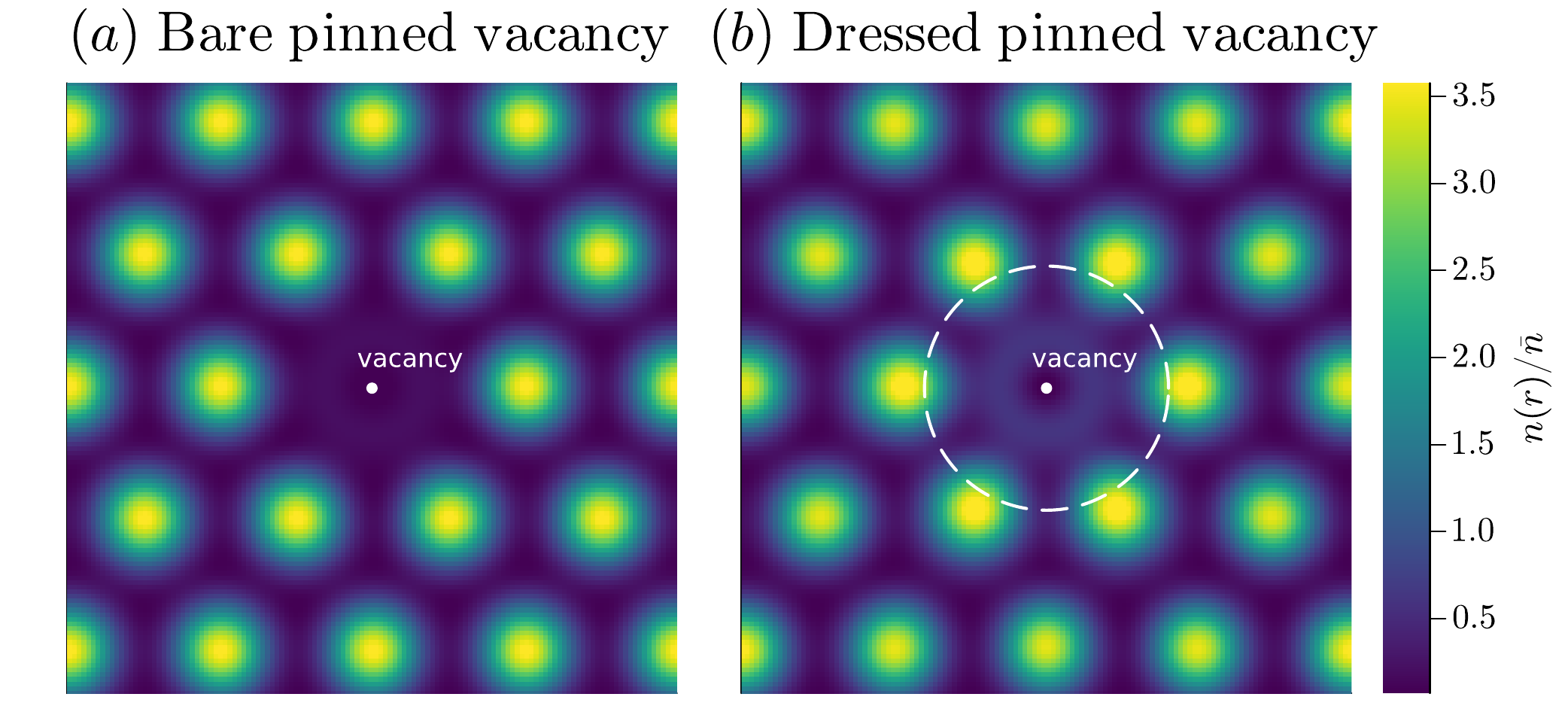}
	\end{subfigure}
	\caption{ \textbf{(Top)} Energies of the commensurate CDW and aWC
compared with those of one-vacancy hCDW and dressed pCDW states with a hole at
$\Ga$, where the hole creation is most energetically competitive. We find that
the hCDW$_\Ga$ is energetically competitive from around $0.4 \times
10^{12}\,\mathrm{cm}^{-2}$. The dressed pCDW$_\Ga$ state performs even better
with noticeable improvements compared to CDW and hCDW$_\Ga$. Range of errors
are given in the ribbon. The same offset function is used. \textbf{(Bottom)}: Visualization of the
single-determinant density for the bare pinned vacancy $D_0 (\{ \r_i\})$
(left) and the dressed pinned vacancy $F_0 (\{ \r_i\}) D_0 (\{ \r_i\})$
(right) for  $A_p = 1.2$ and $\xi_p = 0.8$ in $\sqrt{n_e}=1$ units, which are
the optimized values for the polaron at $n_e = 0.4 \times 10^{12}$cm$^{-2}$.
The dressing enhances electron density on the first shell surrounding the
vacancy. The density is normalized to the average density of the system.}
\label{fig:hCDW}
\end{figure}

From the CDW ansatz, the straighforward route to construct a self-doped state
is to construct the Slater determinant of Bloch states within the lowest band,
but with some momentum sectors omitted while maintaining the universal Jastrow
factor. A natural choice is to have the orbitals near $\k=0$ omitted, which we
refer to the resulting many-body wavefunction as a hole-doped CDW (hCDW$_\Ga$,
where $\Ga$ indicates that the holes are near $\k=0$; see Fig.~\ref{mCDW}). In
this paper, we find that the doped hCDW$_\Ga$ state can be competitive with
the CDW state (see Fig.~\ref{fig:hCDW}) at relevant densities around $0.4
\times 10^{12}$cm$^{-2}$, with the cost in Coulomb energy and the gain in
kinetic energy almost canceling each other.

We can make the doped state be more than competitive by considering a doped
hole screened by WC distortions, which we call a polaron. To construct a
many-body wavefunction for a polaron, we choose localized Wannier
wavefunctions as the basis for the single-particle states in the lowest band,
with the Wannier centers arranged on the triangular WC lattice. The Slater
determinant can then be dressed by two correlation factors. The first is the
conventional Jastrow factor. The second is a new correlation factor between
electron position and orbital index, where the orbital index labels the
position of a hole, i.e., an unoccupied Wannier orbital. This construction
describes a dressed vacancy, or polaronic hole, and we further allow this
polaron to hop and form a plane wave. We refer to the resulting variational
state as the polaron CDW (pCDW).  We find that the energy of the pCDW$_\Ga$
state can have even lower energies compared to hCDW$_\Ga$ near the transition
($n_e \simeq 0.4\times 10^{12}\,\mathrm{cm}^{-2}$), and below the CDW energy
(see Fig.~\ref{fig:hCDW}), indicating that hole-self-doping instability can be
more readily achieved by considering electron density relaxation effects.


\textit{Slater-Jastrow wavefunction for the trial states}---For the
ans{\"a}tze for QFL, commensurate crystals, and hCDW incommensurate crystals,
we construct a Slater-Jastrow type wavefunction:
\begin{equation}
	\Psi(\{ \r_i \}) = \underset{ij}{\textrm{det}}\: [\psi_{\lambda_j}( \r_i) ]J(\{ \r_i\}),
\end{equation}
where $\lambda_j$ is some quantum number associated with the electron such as
crystal position (for crystals) or wavevector (for QFL). The different
crystals will differ in the single-particle orbital structure while the
Jastrow factor will remain universal throughout. 

The gWC state incorporates standard Gaussian wavepacket $\psi_{gWC}(\lvert
\r_i -\mb R_j \rvert) = \exp{-C \lvert \r_i - \mb R_j \rvert^2}$, contrasted
with the aWC state which applies a momentum high-pass filter:
\begin{equation}
	\psi_{aWC}(\lvert \r_i - \mb R_j \rvert) = \int_0^\infty 
\hskip -2mm
\dd k \: k \psi_{gWC}(k) \frac{J_0(k\lvert \r_i - \mb R_j \rvert) }{1+\ee^{-\xi (k-k_F)}},
\end{equation}
where $C$ is the width of the Gaussian orbitals, $k_F$ the high-pass momentum
threshold, $\xi$ the cutoff length  and $\psi_{gWC}(k)$ is the Fourier
transform of the Gaussian wavepacket. A detailed derivation of the orbitals is
given in supplemental material~\footnote{References \cite{GNZF250205383,
BM7715, TC8905, HJ260400113, KK230913121} are relevant in the supplemental
material.}.


For the CDW ansatz, we start from the Mexican-hat dispersion relation 
\begin{equation}
	\epsilon = c_2 k^2 + c_4 k^4,
	\label{eq:disp}
\end{equation}
with $c_2 <0 $ and $c_4>0$. We introduce an auxiliary periodic potential,
which will act as creating orbitals which are centered at Wigner crystal
lattice sites. The auxiliary potential therefore makes the construction to be
a periodic single-particle problem
\begin{equation}
	H_{aux} = c_2 k^2 + c_4 k^4 - 2V_0 \sum_{j} \cos (\mb G_j \cdot \r),
\end{equation}
where $\mb G_j$s with $j=1, 2, 3$ are the three shortest reciprocal lattice vectors of the triangular Wigner crystal~\footnote{While it does not matter due to the symmetry, it is convenient to define $\mb G_1$ and $\mb G_2$ to be 120 degrees apart and $\mb G_3 = -\mb G_1 -\mb G_2$.}. This construction ensures that for larger $V_0$, the orbitals will be more localized near the WC lattice points.

Denoting the eigenstates of this auxiliary Hamiltonian as $\psi_{n\k} (\r)$ and taking the lowest band states $n=1$, the CDW ansatz is by taking all Slater determinant elements to be the Bloch orbitals:
\begin{equation}
	\Psi_{\textrm{CDW}}(\{\r_i \}) = \underset{\k }{\textrm{det}} \left[\psi_{1\k} (\r_i)\right] J(\{\r_i \}).
\end{equation}
The construction of hole states is achieved very easily by removing specific $\k$ orbitals:
\begin{equation}
	\Psi_{\textrm{hCDW}}(\{\r_i \}) = \underset{\k \notin \textrm{holes} }{\textrm{det}} \left[\psi_{1\k} (\r_i)\right] J(\{\r_i \}).
\end{equation}

Finally, the QFL trial states will incorporate the standard form $\psi_{QFL}(\r_j) =
e^{i\k_i \cdot \r_j}$ for a particle occupying the momentum state $\k_i$. To incorporate the Mexican-hat dispersion, we construct the trial states by omitting all $\k$ states which are $\lvert \k \rvert < k_F^{in}$, which will therefore be a ring-like Fermi surface.  

The Jastrow factor is defined in a standard manner where $J(\{ r_i \} ) =
e^{\sum_{i<j} u(\lvert \mathbf r_i - \mathbf r_j \rvert)} $. We use a
combination of short-range and long-range correlators given as $u(\lvert
\mathbf r_i - \mathbf r_j \rvert) = u_{SR} + u_{LR}$. The short-range
correlator is written in real space as
\begin{equation} 
	u_{SR}(\lvert \mathbf r_i - \mathbf r_j \rvert ) = -A\left(1 + \frac{r}{B} + \frac{r^2}{2B^2}\right) e^{-\frac{r}{B}}.  
\end{equation} 
The long-ranged correlator $u_{LR}$ is more effectively written in momentum space and is given as 
\begin{equation}
	u_{LR}(q) = -\frac{A_{LR}}{(q^2 + q_0^2)^{5/4}} e^{-(q/q_c)^4}, 
\end{equation}
motivated by the Gaskell form used in 2DEG. $A$, $B$, and $A_{LR}$ are the
Jastrow VMC parameters. For a generic quartic dispersion relation, the quartic
differentiation indicates that the cusp condition should have $u' = 0$, and we
note that this Jastrow factor satisfies this condition.

For all competing commensurate states, we perform simulations at $n_e=1$ units
and for 64 electrons on a torus. For the incommensurate state, we remove one
electron from the system while maintaining the $n_e = 1$ units for
consistency, which therefore changes the lattice constant of the Wigner
crystal in the simulation. 
The details of the implementation is provided in the supplemental material.
All plotted energies are energies per electron unless stated otherwise. 




\textit{Polaron correlations}---One possible direction to further improve from
the hole picture is to incorporate density relaxation effect, which
effectively simulates a polaron, which has recently been
suggested~\cite{KK230913121} as a possible pathway to self-doping instability
in 2DEG. To incorporate electron density relaxation around a vacancy at a
Wigner crystal site, it is useful to construct the Slater determinant using
real-space Wannier orbitals instead of the Bloch basis~\footnote{We note that
the Bloch functions have arbitrary phases: $\psi_{1\k} \rightarrow
e^{i\theta(\k)} \psi_{1\k}$. We choose the convention that for each Bloch
eigenvector, the wavefunction at $\r=0$ is real and positive. }. The Wannier
orbital centered at site $\mb R$ can be defined as:
\begin{equation}
	w_{\mb R} (\r) =\frac{1}{ \sqrt{N}} \sum_{\k} e^{-i \k \cdot \mb R} \psi_{1\k}(\r).
\end{equation}
After dressing, the single-polaron pCDW ansatz can be written as
\begin{equation}
	\Psi_{\k}^{pCDW} (\{ \r_i \}) = J(\{ \r_i\})  \sum_{\mb R} e^{-i \k \cdot \mb R} F_{\mb R}(\{ \r_i \})D_{\mb R} (\{ \r_i\}),
\end{equation}
where 
\begin{equation}
	D_{\mb R} (\{ \r_i\}) = \underset{a \in \Lambda \setminus \{ \mb R\}}{\textrm{det}} \left[w_{\mb R_a} (\r_i) \right],
\end{equation}
is the pinned defect determinant which omits the orbital at one site $\mb R$, and \begin{equation}
	F_{\mb R}(\{ \r_i \}) = \exp \left[ A_p  \sum_i \exp(-\frac{\lvert \r_i - \mb R \rvert^2}{2 \xi_p^2}) \right],
\end{equation}
is the dressing term which enhances the electron density around this vacancy, creating a screening effect. $A_p$ and $\xi_p$ are additional variational parameters determining the strength and width of the polaron. Using the optimized values for hCDW and CDW at respective densities, we can fix those values and perform a partial optimization fitting $A_p$ and $\xi_p$. This correlation is beyond conventional Jastrow factor since it is a vacancy-conditioned density-relaxation channel unlike the standard translationally invariant two-body Jastrow factor. The detailed construction and simulation details of the pCDW states are discussed in the supplemental material.

\textit{Energetics of commensurate electron crystals}---First, we compare the
energetics between all the commensurate states against the Fermi liquid, as
presented in Figure~\ref{fig:WCCDW}. As stated above, we have verified the
Mexican-hat parameters we have taken are the effective phenomenological band
parameters calibrated to the experimental Lifshitz and crystallization
densities. 

We observe that gWC is energetically very unfavorable, in contrast to the aWC
and CDW states which are competitive to each other. Since aWC and CDW are
constructed in a complementary way, the fact that the two results agree at low
densities is an optimization check, which we indeed verify. The result can
differ in intermediate densities: we find that the CDW state performs better
than the naive high-pass method employed by the aWC energetically.

Deep inside the Brillouin zone, how well does the Bloch orbital resemble the
bare microscopic $\k=0$ state? For $\Ga$ this is particularly useful, as
having additional $\mb G$ components can change the interpretation of holes.
When there is a large mixing between different $\mb G$ states, the Fourier
components will sum to a structured orbital. A simple diagnostic is plotting
the component ratio for the $\mb G = 0$ sector of the $\psi_{10}$ orbital,
which is shown in the inset of Figure~\ref{fig:WCCDW}. We find that the mixing
only becomes serious around at $0.3 \times 10^{12}$cm$^{-2}$, and over this
density one can interpret $\Gamma$-hole as truly a $\k=0$ Bloch
hole~\footnote{The physics is different for $\mb K$ and $\mb M$ sectors as these are
at the Brillouin zone edges. Here, since there are $\mb K$ vectors which
translate $\mb K$/$\mb M$ to their respective degenerate points, there is
guaranteed degenerate mixing between these orbitals. }. 


\textit{Ground state of the system}---We now compare how the hCDW$_\Gamma$ and pCDW$_\Gamma$ states, which are the two most energetically competitive doped states, perform against the commensurate crystal states.  Figure~\ref{fig:hCDW} shows the result. 


We observe that creating holes at $\Gamma$ (hCDW$_\Ga$) is energetically
competitive against the commensurate CDW and aWC states around $0.4 \times
10^{12}$cm$^{-2}$, which is around where signatures of metallic WC was
observed. This indicates that at this particular parameters for the
dispersion, we approach the self-doping instability possibly before the
density becomes too high. 

While hCDW$_\Ga$ is competitive, the inclusion of polaron correlation can
further push down the energies of the doped state. We indeed see this effect
in Figure~\ref{fig:hCDW}, where a single polaron in pCDW$_\Ga$ has noticeably lower
energies compared to hCDW$_\Ga$. While not relevant in ground
state energetics, we observe a similar amount of enhancement for the $\mb K$
and $\mb M$ pCDW states, which we present in the supplemental material.

Therefore, it is the combined effect of the standard optimized
orbital-electron pair Jastrow correlations alongside the vacancy-coupled
relaxation which achieves the competitiveness of the self-doped state. The
renormalized band picture allows for the kinetic energy to compensate the
Coulomb energy disadvantage by creating a hole, while the dressing can
attenuate the Coulomb energy effect substantially and make the one-vacancy
pCDW state energetically favorable over a substantial density window,
indicating a hole-self-doping instability.

\textit{Estimate of the effective hole band parameters}---From the energetics
data, it is now instructive to construct an effective hole band picture:
\begin{equation}
	E_h(\k) = E_0 - 2t_1 \sum_{\textrm{NN}} \cos (\k \cdot \bm{\delta} ) - 2t_2 \sum_{\textrm{NNN}} \cos (\k \cdot \bm\delta' ),
\end{equation}
where $\bm \delta$ and $\bm \delta'$ are the vectors between the
reference WC lattice point to the NN and NNN sites respectively. From the
energy data for the high symmetry points, the coefficients can be extracted by
$t_1 = \frac{E_{\mb K} - E_\Ga }{9}$ and $t_2 =   \frac{E_{\mb M} - E_\Ga }{8}
-\frac{E_{\mb K} - E_\Ga }{9}$.

\begin{figure}
	\centering
	\includegraphics[width=\linewidth]{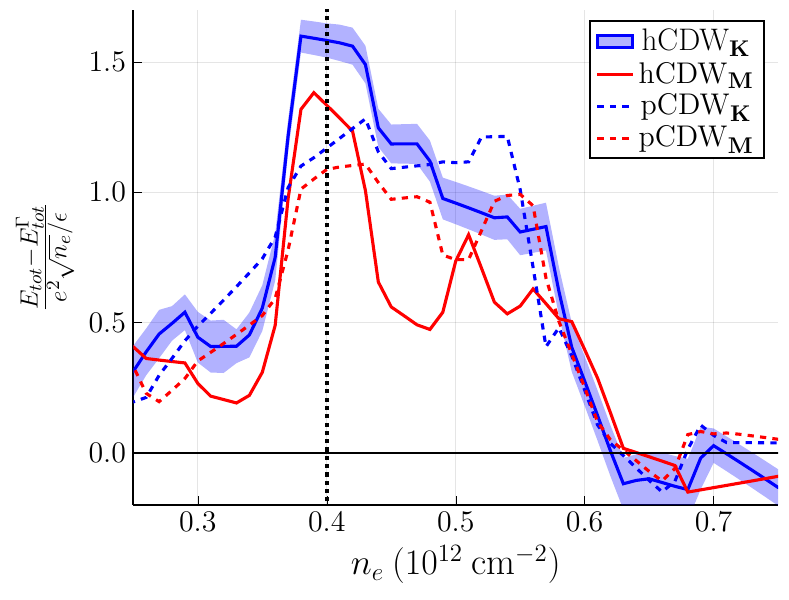}
	\caption{Difference of the \textit{total} energy of the system in units of Coulomb energy for hCDW$_{\mb K}$ and hCDW$_{\mb M}$ states against hCDW$_{\Gamma}$ (in lines) and pCDW$_{\mb K}$ and pCDW$_{\mb M}$ states against pCDW$_{\Gamma}$ (in dashed lines). The energy difference can be used to fit coefficients of the effective hole tight-binding model for the nearest and next-nearest neighbor terms. Errors for hCDW$_{\mb K}$ is displayed; other states have similar size of errors.}
	\label{fig:CDWpol}
\end{figure}

Figure~\ref{fig:CDWpol} shows the energy difference of the whole system
between systems with holes at $\mb K$ and $\mb M$ states compared to their
$\Gamma$ counterpart, for both hCDW and pCDW.
hCDW and pCDW have around similar energy differences, which
indicates a minor renormalization effect of polarons. Picking the relevant
experimental parameter at $0.4 \times 10^{12}$cm$^{-2}$ (dotted vertical
line), $t_1 \approx 0.18 \pm 0.03 E_{c}$ and $t_1 \approx 0.13 \pm 0.03 E_{c}$
for hCDW and pCDW respectively, and for both cases $t_2 \approx 0.00 \pm 0.02
E_{c}$, where $E_c =e^2 \sqrt{n_e}/\epsilon$.
hCDW and pCDW have around similar energy differences, and
its $t_1$ and $t_2$ values are comparable.

While this result ignores the further possibility of hole-hole interaction, we
can still calculate the hole effective mass $m_h^* = \frac{\hbar^2}{3a^2 (t_1
+ 3t_2)}\approx 0.07 m_e$, where $a$ is the WC lattice constant. Naive
dispersion relation retrieves around $m_{h, bare} = \frac{\hbar^2}{2\lvert c_2
\rvert} \approx 0.2 m_e$, hence the hole effective mass is a few times smaller
than the naive band structure result, which provides a good order of magnitude
estimate and agrees with experimental results that the observed hole mass is
smaller than the band effective mass. 

\textit{Discussion and Outlook}---From the numerical point of view, at
Mexican-hat dispersion with coefficients relevant to the experimental point of
interest, the single-particle orbitals of a Wigner crystal should be deformed
significantly. We addressed this problem by introducing an auxiliary periodic
potential to generate an optimized crystalline orbital subspace adapted to the
Mexican-hat dispersion. At low enough densities, the energies are very close
to the naive construction of orbitals where we removed a certain $\k$ space
contributions from the single-particles orbitals directly.

The Wigner crystal states (of various kind) are very good at adapting to the dispersion relation by smearing the orbitals. To be energetically beneficial, the doped state must overcome these deformed CDW states. However, even with the deformed CDW states we find that self-doping is a reasonable possibility. The interaction renormalized band structure allows the kinetic energy to be competitive against potential energy. Combined with the beyond-Jastrow charge relaxation effect, the doped states can be seriously energetically competitive. The effective hole band picture, while more of a crude estimate, does indeed give sensible values for the hopping parameters and support the experimental results of hole mass being smaller than the band calculations.



There are various directions which can be pursued from here. Physically, the concrete way to determine the metallicity will be to investigate the correlation function of the trial Wigner crystal. While more difficult because Berry curvature produces a nontrivial projected-coordinate algebra, its effects can also be investigated, which will play a role in the fine energetics of the various competing states as they are within few percent of Coulomb energy scale. Indeed, this problem has been explored in Bernal bilayer graphene~\cite{JS231007751} and more generally in 2DEG~\cite{JS250722121}. Lastly, while the system size will be smaller due to the $k^4$ dispersion, a neural quantum state analysis of this dispersion will also be complementary to this work, as naturally this will lead to many more interesting crystalline states deep inside the Mexican-hat regime of the phase diagram.

\textit{Note added}--- Related results to the present work will appear in a forthcoming preprint~\cite{AF2608}.

\section*{Acknowledgments}

We would like to thank Tonghang Han, Long Ju, Erez Berg, Taige Wang, Ahmed
Abouelkomsan, Kyung-Su Kim, Zhihuan Dong, Patrick Ledwith, and T. Senthil for
interesting discussions. M.L.K. and X.-G.W. were partially supported by the
Simons Collaboration on Ultra-Quantum Matter, which is a grant from the Simons
Foundation (651446, XGW). Some of the numerical calculations were done on the subMIT
HPC cluster at MIT, and some were done using services provided by the OSG Consortium, which is supported by the National Science Foundation awards 2030508 and 2323298.

\bibliographystyle{ytphys}
\bibliography{all,publst}

\onecolumngrid
\appendix
\newpage

\section{Methodology}\label{MethodologyAppendix}

In this section, we present the details of the construction of the trial wavefunctions and their properties. In the end, we also discuss the variational Monte Carlo routine and the sampling method for the states in consideration.

\subsection{The CDW auxiliary periodic potential model}
We employed a triangular lattice configuration for the Wigner crystal. To ensure that this is well-defined on a torus, we first define a set of lattice points. Defining the lattice vectors as $\mb a_1 = (1, 0)$ and $\mb a_2 = \left(\frac{1}{2}, \frac{\sqrt{3}}{2}\right)$, the lattice points for the Wigner crystal are positioned at $\mb R_{m,n} =  m \mb a_1 + n \mb a_2$ for $0\leq m, n < 8$. Defining the torus directions as $\hat x = (1,0)$ and $ \hat y = (0,1)$, periodic boundary conditions are then enforced with the torus dimensions being $L_1 = 8$ and $L_2 = 4 \sqrt{3}$. All numerics are run at $n_e = 1$ units, and therefore the whole torus is then scaled by the factor $\sqrt{\frac{2}{\sqrt{3}}}$ so that the total area is ensured to be 64.

As outlined in the main letter, the individual orbitals of the commensurate Wigner crystal is obtained by introducing a periodic potential as an auxiliary potential, which will act as creating orbitals which are centered at Wigner crystal lattice sites. The auxiliary potential therefore makes the construction to be a periodic single-particle problem
\begin{equation}
	H_{aux} = c_2 k^2 + c_4 k^4 - 2V_0 \sum_{j} \cos (\mb G_j \cdot \r),
\end{equation}
where $\mb G_j$s with $j=1, 2, 3$ are the three shortest reciprocal lattice vectors of the triangular Wigner crystal.

The form of the Hamiltonian ensures that the eigenstate will have a momentum space label. For a periodic auxiliary potential, the Bloch Hamiltonian at crystal momentum $\k$ has matrix elements
\begin{equation}
	H_{\mb G, \mb G'} = [c_2\lvert \k+\mb G\rvert^2 + c_4\lvert \k+\mb G\rvert^4] \delta_{\mb G, \mb G'} + V_{\mb G -  \mb G'}, 
\end{equation}
and 
\begin{equation}
	V_{\mb G -  \mb G'} = -V_0
\end{equation}
if $\mb G -  \mb G' = \pm \mb G_1, \: \mb G_2 ,\: \mb G_3$ and zero otherwise.

Necessarily, we need to set a cutoff for $\mb G$, which we set $\lvert \mb G \rvert  =3 \lvert \mb b_1 \rvert $. This is safe as the quartic dispersion strongly suppresses states with high $\mb G$. The omitted states begin roughly at the 4th reciprocal shell, so $\lvert \mb G \rvert \sim 4 \lvert \mb b_1 \rvert$ where $\mb b_1$ is one of the reciprocal lattice vectors. The kinetic energy cost of this state is around $\Delta_{\mb G} \sim c_4 \lvert \mb G \rvert^4$, and noting that 
\begin{equation}
	\delta E \sim \frac{\lvert V_0 \rvert^2}{\Delta_{\mb G}} \approx \frac{\lvert V_0 \rvert^2}{256 c_4 \lvert \mb b_1 \rvert^4}.
\end{equation}
Since both $V_0$ and $\lvert \mb b_1 \rvert$ are $O(1)$ values, the $4^4$ suppression is guaranteed to be powerful. Taking this cutoff will work creating at 37-by-37 Hamiltonian, and therefore 37 bands ($n=1, 2, \cdots 37$) throughout the Brillouin zone. Solving this Hamiltonian will yield a eigenvector for each $\k$ point where the elements are the Bloch functions $u_{\mb G}^{(n)} (\k)$, and the Bloch state $\psi_{nk} (\r)$ is recovered as in the main letter:
\begin{equation}
	\psi_{nk} (\r) = \sum_{\mb G} u_{\mb G}^{(n)} (\k) e^{i(\k + \mb G) \cdot \r},
\end{equation}

We note that setting the cutoff to $\lvert \mb G \rvert = 4 \lvert \mb b_1 \rvert$ will give 61 bands in the VMC simulation.

\subsection{Quarter Fermi Liquid}
Defining the torus directions as $\hat x = (1,0)$ and $ \hat y = (0,1)$, this time we set the torus lengths to be equal on both directions. The individual single particle orbitals are now given as plane wave states $e^{i \k_j \cdot \r_i}$. For the ordinary Fermi liquid, we fill from $\k=0$ until 65 states are occupied, ensuring that the 90-degree rotation symmetry so that the Slater determinant remains real\footnote{First, the central $\k =0$ point is degenerate under 4-fold rotation. Second, technically, just enforcing both $\k$ and $-\k$ are occupied is enough, but $L_x = L_y$ gives the additional $C_4$ symmetry.}. In the case of a ring-like Fermi surface, we exclude all $\k$ states which are $k < k_F$, then fill 64 states in the order of small $k$. As before, we ensure the 90-degree rotation symmetry. 

All numerics are run at $n_e = 1$ units, and therefore the torus is set to $L_x = L_y = 8$ in case of ring Fermi surface and $L_x = L_y = \sqrt{65}$ if ordinary Fermi surface. A useful parameter is the universal Fermi wavevector $k_{F0}$ for the spin-polarized Fermi liquid at $n_e = 1$ units, which can be derived as
\begin{equation}
	\frac{1}{(2\pi)^2} \pi k_{F0}^2 = n_e =1 \rightarrow k_{F0} = \sqrt{4\pi} \approx 3.5449,
\end{equation}
and the effective depletion rate can be calculated by computing $r = \left(\frac{k_F}{k_{F0}}\right)^2$. We calculate the optimized energy by selecting the variational parameter spanning all momentum hole sizes $k_F = 0, 1.0, 1.5, 2.0, 2.5, 3.0, 3.5, 4.0, 4.5, 5.0, 6.0$ in $n_e = 1$ units and set the resultant minimized $E_{QFL}$ as the baseline for crystal energy calculation. The $k_F= 0 $ case has 65 electrons as constructed above since $(0,0)$ point has to be accommodated.  This case is only used to determine the Lifshitz transition, and for the purposes of letter, the relevant densities are all after the Lifshitz transition where $k_F >0$, and hence the QFL simulation result presented effectively always is the result from 64 electrons. 

\subsection{Electron correlation Jastrow Factor}
The Jastrow factor used for all our trial states involve a short-range correlator and a long-range correlator. We define the short-ranged Jastrow $u_{SR}$ as
\begin{equation}
	u_{SR} = -A \left( 1 + \frac{r}{B} + \frac{r^2}{2B^2} \right) e^{-\frac{r}{B}},
\end{equation}
which satisfies the cusp condition $u_{SR}' =0 $. $B$ therefore effectively sets the range of the $u_{SR}$\footnote{In reality, this is a smoothly decaying function, and for this Jastrow to decay by $1/e$, the effective length is $r \approx 3.21B$. }. For all purposes, $B$ would be the order of interparticle spacing, lying on the shorter side (see footnote.). 

The gradient and the Laplacian for the Jastrow factors will also be important quantities. For $u_{SR}$, this is given as
\begin{equation}
	\nabla_i u_{SR} = A \frac{r^2}{2B^3} e^{-\frac{r}{B}} \frac{(x, y)}{r}, \quad \nabla_i^2 u_{SR} = A\frac{r(3B-r) }{2B^4} e^{-\frac{r}{B}}.
\end{equation}

The long-ranged Jastrow factor is difficult to define on real-space, and if done so, should be Ewald treated as by definition it will be long-ranged. However, to capture the long tail, it suffices to consider the small $k$-space summation in the reciprocal space. Therefore, we define $u_{LR}$ in reciprocal space:
\begin{equation}
	u_{LR}(q) = -\frac{A_{LR}}{(q^2 + q_0^2)^{5/4}} e^{-(q/q_c)^4}. 
\end{equation}
with $q_0 = \frac{2\pi}{\textrm{max}(L_1, L_2)}$ and $q_c = \frac{2\pi}{B}$. The two regulators are chosen so that $q_0$ regulates the very small $q$-behavior and large distance behavior which would be unstable due to the finite number of electrons in our system, and therefore would have the characteristic size of the smallest possible nonzero $q$ value. The large-$q$ regulator $q_c$ ensures the convergence of the Jastrow factor at very short distances and makes sure that $u_{LR}$ is the long-distance behavior function which controls the distance between $B$ and $L$. 

The overall $u_{LR}(q) \sim \frac{1}{q^{5/2}}$ was chosen to match the tail behavior with the RPA-Gaskell form. The Gaskell Jastrow form for 2DEG is given as
\begin{equation}
	u_{RPA} (q) = -\frac{1}{S_0(q)} + \left( \frac{1}{S_0(q)^2} + \frac{4\pi e^2}{q} \frac{2m}{\hbar^2q^2}  \right)^{1/2}. 
\end{equation}
with $\frac{4\pi e^2}{k} = 2 \frac{2\pi e^2}{k}$ is the Fourier transform for the Coulomb potential. For Wigner crystal, $S_0(k) \sim k^2$ at small $k$ and for the Fermi liquid it is $S_0(k) \sim k$. In a quartic dispersion, we should modify the second term inside the bracket to be $\frac{2\pi e^2}{q} \frac{1}{c_4 q^4}$. Therefore, the leading term at small $k$ is given as $u_{RPA} (q) \sim \frac{1}{q^{5/2}}$. 

The power 4 on the exponent is chosen so that the exponetial regulator does not modify the leading small-$q$ behavior. At small $q$, the exponent will behave as $1- \left(\frac{q}{q_c}\right)^4$, and therefore will not contribute to the leading $\frac{1}{q^{5/2}}$ scaling. 

The real space implementation is by summing over enough amount of discrete $q$ values. Noting that the Jastrow factor in real space is a function of relative distances,
\begin{equation}
	u_{LR} (\r_i - \r_j) = \frac{1}{A_T} \sum_{\mb q} u_{LR} (\mb q) e^{i \bm q \cdot (\r_i - \r_j)}.
\end{equation}
It is convenient to work with the sum of long-distrance Jastrow factors,
\begin{equation}
	U_{LR}(\{\r_i\} ) = \sum_{i<j} u_{LR}(\r_i - \r_j) = \frac{1}{2A_{T}} \sum_{\bm q} u_{LR} (\bm q) \left[\sum_{i} \sum_j e^{i \bm q \cdot (\r_i - \r_j)} - N \right],
\end{equation}
where $A_{T}$ is the area of the simulation area. Writing this in terms of the static structure factor $\rho_{\mb q}$,
\begin{equation}
	U_{LR}(\{\r_i\} )= \frac{1}{2A_{T}} \sum_{\mb q} u_{LR} (\mb q) \left[ \rho_{\mb q} \rho_{-\mb q} - N \right].
\end{equation}
Therefore, the total Jastrow factor can be written as
\begin{equation}
	J(\{\r_i\} ) = e^{U_{SR}(\{\r_i\} )  + U_{LR}(\{\r_i\} )}; U_{SR}(\{\r_i\} )  + U_{LR}(\{\r_i\} ) = -\sum_{i<j} \left[A \left( 1 + \frac{r}{B} + \frac{r^2}{2B^2} \right) e^{-\frac{r}{B}} \right] + \frac{1}{2A_{T}} \sum_{\mb q} u_{LR} (\mb q) \left[ \rho_{\mb q} \rho_{-\mb q} - N \right].
\end{equation}
We comment that the real-space representation of $u_{LR}(r)$ can be written as
\begin{equation}
	u_{LR}(r) = \frac{1}{A_{T}}\sum_{\bm q} u_{LR}(\bm q) \cos (\bm q \cdot \r).
\end{equation}
Therefore, the tail behavior is
\begin{equation}
	u_{LR}(r) - u_{LR}(0) = \frac{1}{A_{T}}\sum_{\bm q} u_{LR}(\bm q) (\cos (\bm q \cdot \r) - 1).
\end{equation}
This is always positive since $\cos (\bm q \cdot \r) - 1$ is always negative, and $u_{LR}(\bm q)$ is also always negative. 

The tail asymptote can easily be obtained just by power counting. Since $u_{LR}(\bm q) \sim \frac{1}{q^{5/2}}$, and $(\cos (\bm q \cdot \r) - 1) \sim q^2$, applying $\int qdq$ leaves the scaling as $u_{LR}(r) - u_{LR}(0) \sim \sqrt{r}$ at intermediate distances. In a finite system, the exponential regulator will ensure that the scale will be cut off at distance comparable to system size and the Jastrow factor is well-defined. One should interpret this scaling as the algebraic behavior at intermediate distances between $B$ and $L_1, L_2$. 

The gradient and the Laplacian for the Jastrow factors will also be important quantities. We first reiterate the gradient and laplacian for the short-range $u_{SR}$:
\begin{equation}
	\nabla_i u_{SR} = A \frac{r^2}{2B^3} e^{-\frac{r}{B}} \frac{(x, y)}{r}, \quad \nabla_i^2 u_{SR} = A\frac{r(3B-r) }{2B^4} e^{-\frac{r}{B}}.
\end{equation}
For the long-ranged Jastrow, once again it is easier to directly work with $U_{LR}(\{\r_i\})$, and
\begin{equation}
	\nabla_i U_{LR}(\{\r_i\} )= \frac{1}{A_{T}} \sum_{\mb q} u_{LR} (\mb q) \left[ i \mb q e^{i \mb q \cdot \r_i} \rho_{-\mb q} \right], \quad \nabla_i^2 U_{LR}(\{\r_i\} )= -\frac{1}{A_{T}} \sum_{\mb q} u_{LR} (\mb q) \left[ q^2 e^{i \mb q \cdot \r_i} \rho_{-\mb q} - q^2 \right], 
\end{equation}
where the 2nd term in the Laplacian ensures that self-summation between particle $i$s are excluded.

For a generic quartic dispersion relation, the quartic differentiation indicates that the cusp condition should have $u' = 0$, and this Jastrow factor satisfies this
condition. For all competing states, we have run the simulations for 64
electrons on a torus if commensurate and 63 if incommensurate unless otherwise noted. We will perform all simulations in $n_e=1$ units, hence
the area of the torus is set to $64$ or $63$.  All values of the variational parameters
are in $n_e=1$ units. The details of the simulation is provided in
later in this supplemental material. 

\subsection{Sampling the Energy}

\subsubsection{Kinetic Energy for Slater-Jastrow wavefunction}
As per VMC protocol, kinetic energy is sampled by taking the local energy. Given an operator, the expectation value is given as
\begin{equation}
	\langle O \rangle = \frac{\int \psi^* O \psi d\mb R}{\int \psi^* \psi d\mb R}= \frac{\int \psi^*  \psi \frac{O \psi}{\psi} d\mb R}{\int \psi^* \psi d\mb R} = \left\langle \frac{O \psi}{\psi} \right\rangle_{MC}.
\end{equation}
where $\mb R$ are the collective coordinates for all particles involved. Therefore, for the kinetic energy terms $k^2$ and $k^4$, the appropriate quantities to sample are
\begin{equation}
	\langle k^2\rangle = -\left\langle \frac{\nabla^2 \psi}{\psi} \right\rangle_{MC}, \quad \langle k^4\rangle = \left\langle \frac{\nabla^4 \psi}{\psi} \right\rangle_{MC}.
\end{equation}
However, there is an alternative sampling for $\langle k^4 \rangle$, since a torus does not have a boundary. Instead of $\frac{\int \psi^* \nabla^4 \psi d\mb R}{\int \psi^* \psi d\mb R}$, one can perform integration by parts twice. The lack of a boundary ensures that there won't be a nontrivial boundary term. Then, an alternative sampling quantity can be written as
\begin{equation}
	\langle k^4 \rangle = \frac{\int \frac{\partial^2 \psi^*}{\psi^*} \frac{\partial^2 \psi}{\psi} \psi^* \psi d\mb R}{\int \psi^* \psi d\mb R} = \left\langle \left \lvert \frac{\nabla^2 \psi}{\psi} \right\rvert^2 \right\rangle_{MC}.
\end{equation}
A sanity check to check this at analytically known systems, such as the Hartree-Fock result of a Fermi liquid. In $n_e = 1$ units, a spin-polarized Fermi liquid has $\langle k^2 \rangle = 2\pi$ and $\langle k^4 \rangle = \frac{4}{3} (2\pi)^2$. Another sanity check is to compare the results for $\langle k^2 \rangle$ between $\left\langle \left \lvert \frac{\nabla \psi}{\psi} \right\rvert^2 \right\rangle_{MC}$ and $-\left \langle \frac{\nabla^2 \psi}{\psi} \right\rangle_{MC}$. One can see that sampling these two quantities converges to the same result. Therefore, for both quadratic and quartic quantities, the essential quantity to sample is the Laplacian of the wavefunction.

For the Slater-Jastrow function given in the form of $\Psi(\{\r\}) = D(\{\r\}) e^{U(\{\r\})}$ with the Jastrow factor $J(\{\r\}) = e^{U(\{\r\})}$, 
\begin{equation}
	\frac{\nabla_i^2 \Psi(\{\r\})}{\Psi(\{\r\})} = \frac{\nabla_i^2 D}{D} + 2\frac{\nabla_i D}{D} \cdot \frac{\nabla_i J}{J} + \frac{\nabla_i^2 J}{J}.
\end{equation}
For each of the terms, the previously calculated results for the gradients and Laplacian will be involved. It is much more useful to update one electron position each time, and if that is the case then
\begin{equation}
	\frac{\nabla_i D}{D} = \sum_1^N \nabla_i \psi_j (\r_i) d_{ji}^{-1}(\{\r_i\}), \quad  \frac{\nabla_i^2 D}{D} = \sum_1^N \nabla_i^2 \psi_j (\r_i) d_{ji}^{-1}(\{\r_i\})
\end{equation}
with $d_{ji}^{-1}$ being the relevant element of the inverse Slater matrix. Because the determinant is linear in row $i$, the corresponding cofactors are independent of $\r_i$. For the total Jastrow $J(\{ \r_i \})$, we comment that
\begin{equation}
	\frac{\nabla_i J(\{\r_i\} )}{J(\{\r_i\} )} = \nabla_i U_{SR} + \nabla_i U_{LR}, \quad \frac{\nabla_i^2 J(\{\r_i\} )}{J(\{\r_i\} )} = \left(\frac{\nabla_i J}{J} \right)^2 + \nabla_i^2 U_{SR} + \nabla_i^2 U_{LR}.
\end{equation}

The calculation procedure for the polaron kinetic energy is more involved than this procedure, which we present at the next appendix in a separate subsection.

\subsubsection{Potential Energy}
Since the Coulomb energy is long-ranged, naively summing over multiple images does not suffice. We implement the Ewald summation method (see for example, Ref.~\cite{GNZF250205383} for details). 

A useful sanity check is to make sure that when the electrons are positioned exactly at the Wigner crystal lattice points, this should give the exact Madelung result~\cite{BM7715}, which in $n_e=1$ units, should give the Coulomb energy expectation value as
\begin{equation}
	\frac{E_{Coul}}{N} = V \frac{e^2 \sqrt{n_e}}{\epsilon}
\end{equation}
with $V \approx -1.9605$. Note that typically this value is given in units of Hartree/$r_s$. A similar sanity check could also be done for the Hartree-Fock result of the Fermi liquid, whose potential per particle is given as $V = - \frac{8}{3\sqrt{\pi}}$. This value is typically given as $-\frac{8}{3\pi r_s} $ Hartrees in the literature.

\subsection{Parameter selection and optimization}
For the selection of the Mexican hat parameter, we take a semi-iterative approach. First, the QFL, gWC, and aWC result samples $\langle k^2 \rangle / n_e$, $\langle k^4 \rangle / n_e^2$, and $\langle \frac{1}{r} \rangle / \sqrt{n_e}$. Therefore, we can scale these results to produce the Fermi liquid energy at arbitary densities:
\begin{equation}
	\frac{E(n_e)}{N} = c_2 n_e \langle k^2 \rangle_{n_e = 1} + c_4 n_e \langle k^4 \rangle_{n_e = 1}+ \frac{e^2 \sqrt{n_e}}{\epsilon} \langle V  \rangle_{n_e = 1} 
\end{equation}
We set the interacting Lifshitz transition to be at $n_e \approx 0.7 \times 10^{12}$cm$^{-2}$.

The CDW ansatz cannot be scaled as readily, but it can be scaled almost freely by using $n_e=1$ units. Since $c_2$, $c_4$, and $V_0$ all have different units however, this means to maintain $n_e=1$, the simulation values for $c_2$ and $c_4$ must be scaled by the $n_{diag}$ parameter. We converge on the values of $c_2$ and $c_4$ so that the CDW-QFL transition occurs around at $0.4-0.5 \times 10^{12}$cm$^{-2}$.

For the commensurate state, the variational parameters are the three involved in the Jastrow factor ($A$, $B$, and $A_{LR}$), alongside the auxiliary potential strength $V_0$ and diagonalization density $n_{diag}$ which we use for the dispersion relation to construct the Wannier orbitals. 

We perform a grid scan of these parameters with ranges for $A$ being 0.5, 1, 1.5, 2, 3, 4, 6, 8, 10, 12, $B$ being 0.1, 0.15, 0.2, 0.25, 0.3, 0.4, $A_{LR}$ being 0, 1, 2, 5, 8, 10, 15 (so far all in $\sqrt{n_e} = 1$ units), $V_0$ being 0.2, 0.4, 0.8, 1.5, 2.0, 4.0, 10.0, 20.0 meV
, and $n_{diag}$ being 0.2, 0.3, 0.4, 0.5, 0.55, 0.6, 0.7, 0.8, 1.0, 1.2 
$\times 10^{12}$ cm$^{-2}$.  
These parameters proved to be enough to scan the whole range of densities for the given parameters of $c_2$ and $c_4$ and relevant densities. After the grid scan with the approximate ranges of the optimized parameters given, one can then further focus on the vicinity of the optimized parameters to collect the appropriate amount of data needed.

For the incommensurate case, there is a further introduction of polaron parameters which are the amplitude $A_{pol}$ and size $\xi_{pol}$. Instead of doing a full grid search, we proceed by extracting the fully optimized parameters for the commensurate Wigner crystal and hCDW states at certain physical density points between $0.05$ to $0.8 \times 10^{12}$ cm$^{-2}$ . For the five parameters shared, we only perform further optimizations after setting the five parameters to the already optimized parameters via a grid search of $A_{pol}$ and $\xi_{pol}$. The range of $A_{pol}$ is 0 (which will be the bare Bloch hole), 0.2, 0.4, 0.6, 0.8, 1.0, 1.2, 1.4, 1.6, 1.8, and for $\xi_{pol}$ the range is 0.50, 0.70, 0.85, 1.00, 1.20, 1.40, 1.70, 2.00. A useful guidance for $\xi_{pol}$ is that for a good polaron, the size of the polaron should approximately be the size of the Wigner crystal lattice separation.

\subsection{VMC Routine and Sampling}
As per standard VMC protocol, we sample the probability distribution $\lvert \Psi(\{ \r_i \}) \rvert^2$ with the Metropolis-Hastings algorithm. For each run, we equilibriate for $10^5$ steps, with each step moving one particle for a distance of $0.03 L$ in case of the Fermi liquid and $0.03 L_2$ in case of the Wigner crystal. 

To avoid autocorrelation effects, we only sample the local energy and potential every 100 steps so that every particle would have had a reasonable chance to be updated. We find that this method converges much quicker compared to naively sampling every step. We take $1\times10^5$ samples for every realization of the wavefunction. Each given variational parameter is sampled 5 times and their averages are taken individually for $\langle k^2 \rangle$, $\langle k^4 \rangle$, and $\langle V \rangle$. The errors are calculated using the standard errors for $\langle k^2 \rangle$, $\langle k^4 \rangle$, and $\langle V \rangle$ samples and propagating it for the total energy.

\section{Construction of the Polaron ansatz}
In the main text, we overviewed the concept of the polaron ansatz and the construction of its Bloch state. We present this procedure more in detail in this appendix.

One useful approach is to allow local density relaxation around the hole. This corresponds to creating a polaron, and in the Jastrow factor language could be understood as incorporating electron-many body correlation effect which goes beyond typical two-electron correlation function. A similar pathway has been suggested in Ref.~\cite{KK230913121}, albeit in ordinary 2DEG and an analytic expansion from the low density (high $r_s$) limit. 

In general, creating correlations in this picture is not tractable, as the correlation acts on a pinned vacancy. Creating a coherent hopping of multiple pinned vacancies then requires selecting the vacancies, which implies combinatorial scaling. However, in the case of Wigner crystals, the periodic nature of the crystals allow to easily transition into momentum space picture as well, and therefore a coherent Bloch state can be well-defined with reasonable computational load at least for a single polaron.

Let us first define the pinned vacancy Slater determinant
\begin{equation}
	D_{\mb R} (\{ \r_i\}) = \underset{a \in \Lambda \setminus \{ \mb R\}}{\textrm{det}} \left[w_{\mb R_a} (\r_i) \right],
\end{equation}
which omits the orbital at one site $\mb R$. In second quantized notation this state corresponds to $\vert \Phi_{\mb R} \rangle = d_{\mb R} \vert \Phi_{com} \rangle$, with $d_{\mb R} $ being the Fourier transform of the annihilation operator.

We now create a polaron by dressing this vacancy with a one-body factor
\begin{equation}
	F_{\mb R}(\{ \r_i \}) = \exp \left[ A_p  \sum_i \exp(-\frac{\lvert \r_i - \mb R \rvert^2}{2 \xi_p^2}) \right],
\end{equation}
which enhances the electron density around this vacancy using a simple Gaussian model, creating a screening effect. The dressed pinned state is therefore
\begin{equation}
	\Psi_{\mb R}(\{ \r_i\}) = F_{\mb R}(\{ \r_i \})D_{\mb R} (\{ \r_i\}). 
\end{equation}

The final wavefunction we will use is the full mobile polaron wavefunction, which is the Bloch state of the polaron (pCDW), with the attached Jastrow factor:
\begin{equation}
	\Psi_\k^{pCDW} (\{ \r_i \}) = J(\{ \r_i\})  \sum_{\mb R} e^{-i \k \cdot \mb R} F_{\mb R}(\{ \r_i \})D_{\mb R} (\{ \r_i\}). 
\end{equation}


\subsection{Kinetic Energy for Polaron dressed states}
The methodology presented in the previous section only applies for ordinary Slater-Jastrow wavefunction has to be modified for the polaronic pCDW states. Starting from the pCDW wavefunction stated:
\begin{equation}
	\Psi_\k^{pCDW} (\{ \r_i \}) = J(\{ \r_i\})  \sum_{\mb R} e^{-i \k \cdot \mb R} F_{\mb R}(\{ \r_i \})D_{\mb R} (\{ \r_i\}),
\end{equation}
a useful trick is that this wavefunction can still be written as a product of Jastrow function and a single determinant. Defining the extended matrix $M(\{ \r_i \})$ as
\begin{equation}
	M(\{ \r_i \}) = \begin{pmatrix}
		w_1 (\r_1) & w_2 (\r_1) & \cdots & w_N (\r_1) \\
		w_1 (\r_2) & w_2 (\r_2) & \cdots & w_N (\r_2) \\
		\vdots & & & \vdots \\
		w_1 (\r_{N-1}) & w_2 (\r_{N-1}) & \cdots & w_N (\r_{N-1}) \\
		V_1 (\{ \r_i\}) & V_2 (\{ \r_i\}) & \cdots & V_N (\{ \r_i\})
	\end{pmatrix},
\end{equation}
with $V_{\mb R} (\{ \r_i\}) = e^{-i \k \cdot \mb R} F_{\mb R}$, then
\begin{equation}
	\Psi_\k^{pCDW} (\{ \r_i \}) = J(\{ \r_i\}) \: \textrm{det} \: M(\{ \r_i \})
\end{equation}
after setting the conventions correctly. Our conventions imply that $D_{\mb R} (\{ \r_i\}) = (-1)^{N+R} \mathrm{det} A_{\hat {\mb R}}(\{ \r_i\}) $, where $A_{\hat {\mb R}}(\{ \r_i\}) $ is the square matrix with column $R$ deleted, and $R$ is the index number for $\mb R$. This is by itself consistent with the Bloch hole picture,  defining the filled Wannier state as $(d_1^\dagger d_2^\dagger \cdots d_R^\dagger \cdots d_N^\dagger) \vert 0 \rangle$,
\begin{equation}
	d_R (d_1^\dagger d_2^\dagger \cdots d_R^\dagger \cdots d_N^\dagger) \vert 0 \rangle = (-1)^{R-1} (d_1^\dagger d_2^\dagger \cdots d_{R-1}^\dagger d_{R+1}^\dagger \cdots d_N^\dagger)\vert 0 \rangle 
\end{equation}
and therefore
\begin{equation}
	\Psi_{d_R}(\{ \r_i\}) =  (-1)^{R-1}  \mathrm{det} A_{\hat {\mb R}}(\{ \r_i\}).
\end{equation}
Taking the ratio between $\frac{D_{\mb R} (\{ \r_i\})}{\Psi_{d_R}(\{ \r_i\}) } = (-1)^{N+1} $, which is an overall sign and independent of $R$. Hence, there are no relative sign errors between vacancy positions. 

Moving one electron position $\r_i$ now corresponds to changing two rows, one at the corresponding row of matrix $M$ and the last row containing the other factors. We will write each row of $M$ as $a_i^T (\{ \r_i\}) $  and the last row as $V^T$, and write the $e_i$ basis vector as the $N$-by-1 matrix with $i$th element being 1 and all others and $e_{aug} = e_N$. Then
\begin{equation}
	\partial_{x_i} M = e_i a_{i, x}^T + e_{aug} V_x^T
\end{equation}
where $a_{i, x}^T$ and $V_x^T$ are the $x$-derivatives for each row. Then, 
\begin{equation}
	\partial_{x_i} M = U V_x^T
\end{equation}
with $U = (e_i, e_{aug})$ and $V_x^T = \begin{pmatrix} a_{i,x}^T \\ V_x^T \end{pmatrix}$
with a similar formula for $y$-derivatives.

From this point, Jacobi's determinant formula reduces the calculation to 2-by-2 matrix. Using the formula,
\begin{equation}
	\frac{\partial_\alpha \mathrm{det} M}{\mathrm{det} M} = \mathrm{Tr} (M^{-1} \partial_\alpha M) = \mathrm{Tr} (V_x^T M^{-1} U) 
\end{equation}
using the cyclic property of the trace. The matrix $C_\alpha = V_x^T M^{-1} U$ is 2-by-2, and therefore
\begin{equation}
	\frac{\partial_{i, \alpha} \:  \mathrm{det} M }{ \mathrm{det} M} = \mathrm{Tr} (C_\alpha).
\end{equation}
The Laplacian result can be obtained by differentiating once more:
\begin{align}
	\frac{\partial_\alpha^2 \mathrm{det} M}{\mathrm{det} M} &= \mathrm{Tr} (M^{-1} \partial_\alpha^2 M) + [\mathrm{Tr} (M^{-1} \partial_\alpha M)]^2 - \mathrm{Tr}[ (M^{-1} \partial_\alpha M)^2] \\
	&=\mathrm{Tr} (V_{\alpha \alpha}^T M^{-1} U) + [\mathrm{Tr} \: C_\alpha]^2 - \mathrm{Tr}[ (C_\alpha)^2]
\end{align}
and the Laplacian is the sum of $\alpha = x, y$. This is effectively the rank-two analogue of the familiar Slater inverse-matrix formula.

We now specify how to calculate individual derivatives $a_{i, x}^T$ and $V_x^T$. For the case of $a_{i, x}^T$, we use the Bloch basis as this yields a simple form for the derivatives
\begin{equation}
	\psi_{1\k_{\beta} } (\r_i)  = \frac{1}{\sqrt{N}} \sum_p c_{\beta p} e^{i \mb p \cdot \r_i}; \quad \partial_x \psi_{1\k_{\beta} } (\r_i)  = \frac{1}{\sqrt{N}} \sum_p ip_x c_{\beta p} e^{i \mb p \cdot \r_i}; \quad \nabla^2 \psi_{1\k_{\beta} } (\r_i)  = -\frac{1}{\sqrt{N}} \sum_p p^2 c_{\beta p} e^{i \mb p \cdot \r_i},
\end{equation}
and we Fourier transform back. 

For $V_x^T$, it is convenient to write the polaronic dress factor as
\begin{equation}
	F_{\mb R}(\{ \r_i \}) = e^{\sum_i f_{\mb R}(\r_i)},
\end{equation}
where
\begin{equation}
	f_{\mb R}(\r_i) = A_p  \exp(-\frac{\lvert \r_i - \mb R \rvert^2}{2 \xi_p^2}),
\end{equation}
and we can now use a similar trick one used for the Jastrow factor that
\begin{equation}
	\partial_{x_i} V_{\mb R} = V_{\mb R} \partial_{x_i} f_{\mb R}(\r_i); \quad \nabla_i^2 V_{\mb R} = V_{\mb R} [(\nabla_i f_{\mb R}(\r_i))^2 + \nabla_i^2 f_{\mb R}(\r_i)]. 
\end{equation}

Combining the Jastrow factor is straightforward now, with replacing $\frac{\nabla_i^2 D}{D}$ and $\frac{\nabla_i D}{D}$ with $\frac{\nabla_i^2 \mathrm{det} M}{\mathrm{det} M}$ and $\frac{\nabla_i \mathrm{det} M}{\mathrm{det} M}$ from the previous Slater-Jastrow discussion. 

\subsection{Improvement achieved by pCDW compared to hCDW states}

\begin{figure}[ht!]
	\centering
	\includegraphics[width=0.5\linewidth]{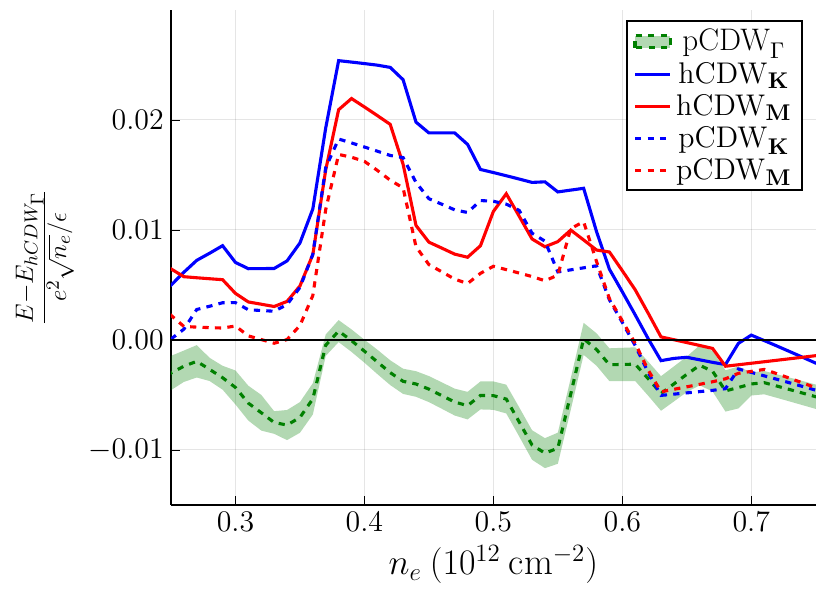}
	\caption{Energies of the doped CDW ans{\"a}tze per electron compared against hCDW$_\Ga$. The effect of polarons is substantial, showing between 0.5-1\% of Coulomb energy per particle for all symmetry point holes compared to their hCDW counterparts.}
	\label{fig:hCDWdiff}
\end{figure}

In the main letter, we presented the energetics result for pCDW and hCDW with a hole at $\Ga$. It is instructive to see how much energy gain per particle the polaron dressing induces, including for pCDW$_{\mb K}$ and pCDW$_{\mb M}$ states against their hCDW counterparts. Figure~\ref{fig:hCDWdiff} shows the difference of all doped states at symmetry points from between hCDW$_{\Gamma}$.

The results show that the polaron dressing achieves a substantial improvement in general compared to their hCDW counterpart, around 0.5-1\% of Coulomb energy per particle. This range is consistent for all symmetry points considered, and shows the effectiveness of the polaron dressing even with the partial optimization technique we have employed. This substantial improvement can be crucial to induce self-doping.

\section{The Annular Wigner crystals with momentum space depletion}\label{app:data}
In this section, we provide additional results for the ansatz where the single-particle orbitals are directly constructed by removing $\k$ space contributions near $\k= 0$. This ansatz will serve as a useful sanity check at low densities and also will provide additional understanding for the shape of the single-particle orbitals for Wigner crystals subjected to Mexican hat potential.

\subsection{Single particle orbitals with removed contributions near $\k = 0$}
A classic Wigner crystal ansatz in real space is in Slater-Jastrow form with Gaussian single particle orbitals, which can be written as
\begin{equation}
	\Psi_{gWC}(\{ \mathbf r\}) = \textrm{det} [\psi_i(\mathbf r_j)] \exp\left(\sum_{i<j} u(\lvert \mathbf r_i - \mathbf r_j \rvert )\right).
\end{equation}
where $\{ r \}$ is the collective electron coordinates, $\mathbf r_i$ is the electron coordinate for electron $i$, and $u(\lvert \mathbf r_i - \mathbf r_j \rvert)$ is the Jastrow factor which incorporates electron correlation. The canonical choice~\cite{TC8905} for single particle orbitals is of form $\psi_i(\mathbf r_j) = \exp(-C \lvert R_i - r_j \rvert^2)$, which is known to provide good energies for 2DEG systems, where $R_I$ is the position of the Wigner crystal lattice points and $r_j$ is the position of the electrons. In 2DEG, it is also known that the triangular Wigner crystal has the lowest energy~\cite{BM7715} is the most energetically stable configuration, albeit being quite close to the square lattice counterpart.

We now consider the situation where we directly remove the contributions from small $\k$-states by applying a momentum space filter for the single particle orbitals. This would correspond to states near $\k$ near 0 not being involved in the ansatz. As our single-particle wavefunction is conveniently in the Gaussian form, we can Fourier transform the single-particle wavefunction and remove the near-zero components. Performing the Fourier transform of $\psi_i(\mathbf r_j)$, we retrieve
\begin{equation}
	\psi_i(\k_j) = e^{-i \k \cdot  \mathbf R_i} \frac{\pi}{C} \exp\left( -\frac{k_j^2}{4C}  \right)
\end{equation}
Naively, we can then construct the new single-particle wavefunction by integration the Fourier component from a momentum cutoff $k_F$ to infinity. After performing the angular integration, this yields a Bessel function of the first kind of order zero $J_0 (k \rho)$:
\begin{equation}
	\psi_{aWC}' (\rho) =  \int_{0}^\infty dk \: k e^{-\frac{k^2}{4C}} J_0(k\rho) \theta( k-k_F),
\end{equation}
where $\rho = \lvert R_i - r_j \rvert$.

This involves a long-ranged algebraic tail at long distances with $\psi_{aWC} (\rho) \propto \frac{\cos (k_F \rho + \delta)}{\rho^{3/2}}$, coming from the sharp cutoff. In order to control this, instead of the sharp Heaviside cutoff, we use a Fermi-Dirac type smearing function:
\begin{equation}
	\psi_{aWC} (\rho) =  \int_{0}^\infty dk \: k e^{-\frac{k^2}{4C}} J_0(k\rho) \frac{1}{1+e^{-\xi (k-k_F)}},
\end{equation}
where $\xi$ will also function as a variational parameter. By using this form, we can control the size of the tail as $\psi_{aWC} (\rho) \propto e^{-\pi \rho/\xi} \cos (k_F \rho + \delta)$ at long distances, alongside providing a smooth variational cutoff.

Therefore, a valid ansatz for Wigner crystal with momentum cutoff $k_F$ will be
\begin{equation}
	\Psi_{aWC}(\{\mathbf r\}) = \textrm{det} [\psi_{aWC}(\rho)] \exp\left(\sum_{i<j} u(\lvert \mathbf r_i - \mathbf r_j \rvert )\right).
\end{equation}

\begin{figure}[ht!]
	\centering
	\includegraphics[width=\linewidth]{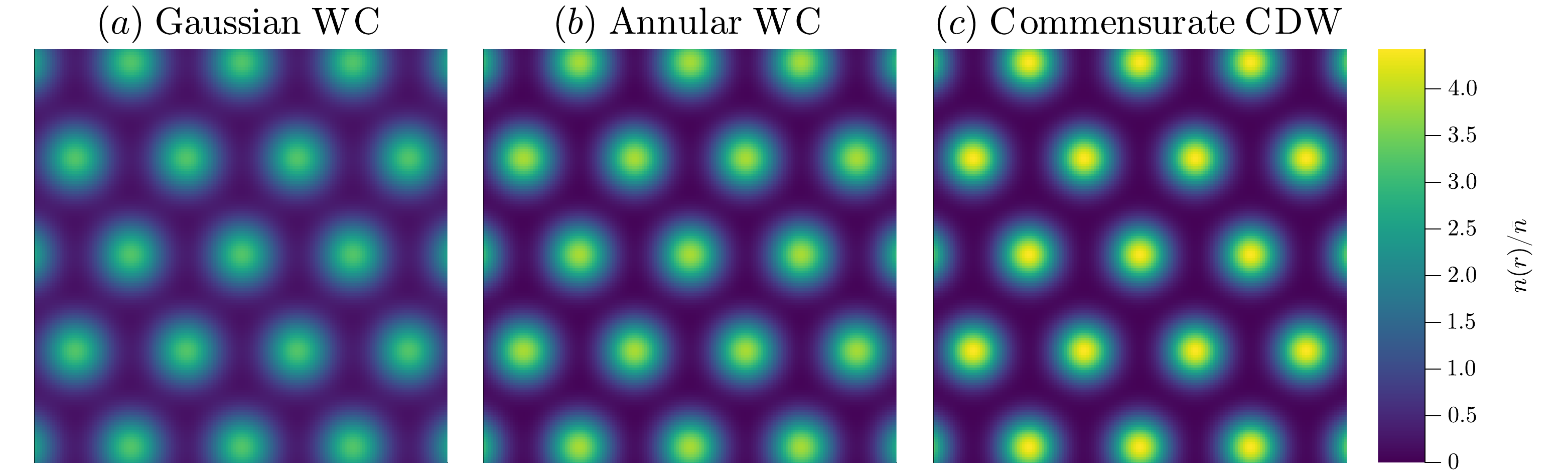}
	\caption{Visualization of the single-determinant density for the gWC (a), aWC (b), and commensurate CDW (c) for the optimized state at $n_e = 0.4 \times 10^{12} \: \textrm{cm}^{-2}$. This corresponds to $C=5$ for gWC, and $\xi=1$, $C=3$, and $k_F = 4.0$ for aWC. The density is normalized to the maximum density of the system (which are exactly at the crystal lattice points).
	}
	\label{fig:visual}
\end{figure}

We note that this will approach the Gaussian Wigner crystal ansatz in the case when $k_F = 0, \xi \rightarrow \infty$. Nonzero $k_F$ being preferable implies that the single-particle orbital has been deformed to remove momentum space contributions from $\k=0$ to $\lvert \k \rvert = k_F$ to accommodate the mexican hat potential, especially when the density is low enough in case of nonzero $k_F$. Figure~\ref{fig:visual} contrasts the single-particle Slater determinant wavefunction of the Wigner crystal normalized to the maximum density for the annular Wigner crystal and ordinary Wigner crystal for 64 electrons.

\end{document}